\documentclass[12pt]{article}

\usepackage{graphicx,amssymb,url,mathrsfs, amsmath}
\usepackage{eucal}
\usepackage{wrapfig}
\usepackage{boxedminipage}
\usepackage{setspace}
\usepackage{subfigure}

\def\IR{{\hbox{{\rm I}\kern-.2em\hbox{\rm R}}}}
\def\IB{{\hbox{{\rm I}\kern-.2em\hbox{\rm B}}}}
\def\IN{{\hbox{{\rm I}\kern-.2em\hbox{\rm N}}}}
\def\IC{\,\,{\hbox{{\rm I}\kern-.59em\hbox{\bf C}}}}
\def\IZ{{\hbox{{\rm Z}\kern-.4em\hbox{\rm Z}}}}
\def\IP{{\hbox{{\rm I}\kern-.2em\hbox{\rm P}}}}
\def\IH{{\hbox{{\rm I}\kern-.4em\hbox{\rm H}}}}
\def\ID{{\hbox{{\rm I}\kern-.2em\hbox{\rm D}}}}

\def\be{\begin{equation}}
\def\ee{\end{equation}}
\def\ba{\begin{eqnarray}}
\def\ea{\end{eqnarray}}

\def\ea{{\it et al}. }

\begin{document}

\begin{titlepage}

\vspace{0.5in}

\begin{center}
{\large \bf Anomalous Chiral Superfluidity}\\
\vspace{10mm}
Michael Lublinsky$^{a,b}$ and Ismail Zahed$^{a}$\\
\vspace{5mm}
$^a$  {Department of Physics and Astronomy, Stony Brook University, Stony Brook NY 11794}\\
$^b$ {\small Physics Department, Ben-Gurion University, Beer Sheva 84105, Israel}\\
          \vspace{10mm}
\end{center}
\begin{abstract}
We discuss both the anomalous Cartan currents and the energy-momentum tensor in a left chiral
theory with flavor anomalies as an effective theory for flavored chiral phonons in  a chiral superfluid with
the gauged Wess-Zumino-Witten term.  In the mean-field (leading tadpole) approximation the anomalous 
Cartan currents and the energy  momentum tensor take the form of constitutive currents in the chiral
superfluid state.  The pertinence of higher order corrections and the Adler-Bardeen theorem is briefly noted.
 \end{abstract}
\end{titlepage}

\renewcommand{\thefootnote}{\arabic{footnote}}
\setcounter{footnote}{0}



\section{Introduction}

Quantum anomalies play an important role at low energy where they condition the character of
the anomalous decays such as $\pi^0\rightarrow \gamma\gamma$, $\gamma\rightarrow \pi\pi\pi$
$....$ decays. They translate the high energy content of a gauge theory in a way that is protected 
from radiative corrections and non-perturbative phenomena. A salient example of these anomalies
is the Wess-Zumino-Witten term and its relevance to low energy chiral dynamics whether in meson
or meson-hadron physics. Its form follows solely from geometry and gauge invariance. The importance
of anomalies in dense QCD with an emphasis on the superfluid phase was noted in~\cite{CFL}.
Recently, similar anomalies have surfaced in holographic QCD at finite density in the context
of a hydrodynamical analysis~\cite{ADS}.

Hydrodynamics is an effective description of long-wavelength physics that encapsulates the 
constraints  of general conservation laws and symmetries.  It describes the flow of the energy momentum
tensor and charged currents beyond the realm of perturbation theory. Although phenomenological in
character, with flow and dissipation encoded in terms of transport parameters, hydrodynamics has been
successful in describing many phenomena ranging from the fundamental such as ultrarelativistic heavy
ion collisions to more mundane such as water flow.

An interesting question regarding the role of anomalies in the hydrodynamical set up was recently
raised in~\cite{ADS,Torabian,SON1}, with a critical discussion in~\cite{ANTON}.
Using arguments based on triangle anomalies and thermodynamics 
they were led to an amendment of the constitutive currents~\cite{SON1}. 
Specifically, they have found that the 
constitutive but anomalous currents support additional terms as dictated by global anomalies that 
involve new transport parameters. We show in this letter, that the amendments of the constitutive currents
are in general expected from the Wess-Zumino-Witten action in the superfluid state~\cite{CFL}.  Here,
we derive these amendments in the mean-field (leading tadpole) approximation. We also note the
relevance of the Adler-Bardeen non-renormalization theorem at the quantum level.

\section{Anomalous Current}

Most of our discussion of anomalies will be centered on the QCD flavor anomalies 
and their transcription to the constitutive flavor currents in a superfluid.  A specific 
example would be QCD at high fermion density and low temperature in the superfluid
CFL phase with global $SU(N=3)_{c+L+R}$ symmetry~\cite{CFL}.  Another, would be
just low temperature QCD with global $SU(N)_{L+R}$ symmetry, or the standard model
in a superfluid state with only left handed fermions at low temperature.

For simplicity 
and notational convenience, we consider generically a left-handed QCD like theory
in the spontaneously broken phase with global $SU(N)$ symmetry in the low
temperature regime.  In the presence of external flavor gauge fields $A_L$, the
flavor currents are anomalous

\be
\nabla_L J_L^a
=2C\,{\rm Tr}\left(T^a\,(dA_L dA_L+\frac 12 dA_L^3)\right)
\label{LEFT}
\ee
with $C=N_c/48\pi^2$, $\nabla_L=d+A_L$   and $A_L=iT^a A_L^a$.  The form notation for the anomaly 
part will be assumed. 
(\ref{LEFT}) holds also for the right current with the substitution $L\rightarrow R$
and $C\rightarrow -C$. Throughout and for simplicity, we will only discuss a chiral theory with left external 
flavor fields $A_L$ in the Cartan  subalgebra of SU(N),  for which (\ref{LEFT}) abelianizes

\be
dJ_L^a=2C\,{\rm Tr}(T^a\,dA_L\,dA_L)
\label{LEFTX}
\ee
The external flavor gauge fields $A_L$ are in general fixed vector sources that play the role of left
 handed
$\rho$ and $a1$ vector fields.  They could be made dynamical by identifying them with the non-Goldstone and
vector excitations of the theory, or held fixed by identifying them with pertinent external gauge fields.

The current $J_L$ in (\ref{LEFT})  or more precisely its expectation value to be defined later, 
can be defined using Witten's anomalous effective Lagrangian in
the superfluid state. Minimally,

\be
\Gamma[U,A_L]=-F^2\int_4\,{\rm Tr}(L+A_L)^2+\Gamma_{WZ}[U,A_L]
\label{WITTEN}
\ee
with $\Gamma_{WZ}$ the gauged Wess-Zumino-Witten effective action~\cite{WITTEN}. 
Here $L=dU\,U^{-1}$ is the left Maurer-Cartan 1-form with zero 2-form (field strength) $dL-L^2=0$.
  For the left Cartan gauge-fields $A_L$, the Wess-Zumino-Witten contribution simplifies

\be
\Gamma_{WZ}[U,A_L]=-\frac {i}5\,C\int_5\,{\rm Tr}(L^5)-iC\int_4\,{\rm Tr}\left(A_LL^3\right)
-\frac i2 C \int_4\,{\rm Tr}(A_LL)^2-i2C\int_4\,{\rm Tr}\left(A_L\,dA_L\,L\right)
\label{WITTENX}
\ee
The effects of temperature
are understood in (\ref{WITTEN}) through a compactification of the time direction.  In general, they 
will lead to the breaking of Lorentz symmetry.  In the mean-field or  tadpole approximation they do not.
(\ref{WITTENX}) obeys the Wess-Zumino consistency condition  for the left anomalous current.

We observe that the Lagrange densities in 
(\ref{WITTEN}) and (\ref{WITTENX}) are invariant under a rigid left Cartan transformation of the
chiral field $U\rightarrow e^{\epsilon_L}U$ for any Cartan valued $A_L$. The associated conserved global 
Cartan current follow through the standard Noether construction, $\tilde{J}_L=\delta \Gamma/\delta d\epsilon_L$

\be
\tilde{J}_L^a=-2iF^2\,{\rm Tr}T^a(L+A_L)-C{\rm Tr}T^a(L+A_L)dL-2C\,{\rm Tr}T^a(A_LdA_L)
\label{CUR2}
\ee
satisfying $d\tilde{J}_L=0$.
We have suppressed the Lorentz indices  for
convenience, and assumed the $L$'s in the Cartan algebra only after the variation of $\Gamma$.  This
is enough for the mean-field analysis to follow.
  The conservation of the Cartan current (\ref{CUR2}) is also the Euler-Lagrange equation
for the chiral field $U$.  For $A_L=0$, $d\tilde{J}_L=0$  together with $dL=L^2$ reduce to the equations of
motion for the chiral field  $U$ with the  Wess-Zumino-Witten contribution.  (\ref{CUR2}) can be rewritten as

\be
\tilde{J}_L^a=-2iF^2\,{\rm Tr}T^a(L+A_L)-C{\rm Tr}T^a(L+A_L)d(L+A_L)+C{\rm Tr}T^a(L+A_L)\,dA_L-2C\,{\rm Tr}T^a(A_LdA_L)
\label{CUR2X}
\ee
with all contributions of the type $L+A_L$ invariant under a local left Cartan transformation
of both $U$ and $A_L$, with the exception of the last Chern-Simons contribution.  We note that
(\ref{CUR2X}) shifts under  $\tilde{J}_L\rightarrow \tilde{J}_L+ d(A_LL)$, without affecting the 
equations of motion since $d^2(A_LL)=0$ is a closed 4-form. This shift amounts to a gauge-variant 
surface term in~(\ref{WITTEN}) which upsets the Wess-Zumino consistency condition for the 
left current. The other possible quadratic shifts $d(L^2)$ and $d(A_L^2)$ are zero for the Cartan labels.

With this in mind, we now recall that the anomalous but gauge invariant left current $J_L$ ties with the 
conserved but gauge variant Noether current $\tilde{J}_L$ through the Chern-Simons density $K$,

\be
\tilde{J}_L^a=J_L^a-K^a=J_L^a-2C\,{\rm Tr}T^a(A_LdA_L)
\label{SIMONS}
\ee
Since $d\tilde{J}_L=0$ by the equations of motion,  the minimal Noether current (\ref{CUR2X}) yields

\be
d\,J_L^a=-\frac{C}{8}\,d^{abc}F_L^bF_L^c
\label{ANO}
\ee
which is (\ref{LEFTX}) for the Cartan charges.

\section{Energy-Momentum Tensor}

The energy momentum tensor associated to (\ref{WITTEN}) follows canonically.  Straightforward
calculations yield

\be
\tilde{T}_{\nu\lambda}=T_{\nu\lambda}-{\rm Tr}\left(A_{L\lambda}\frac{\delta \Gamma}{\delta A^\nu_{L}}\right)
\label{T1}
\ee
with the symmetric energy momentum

\be
T_{\nu\lambda}=-2F^2{\rm Tr}(L+A_L)_\nu(L+A_L)_\lambda+g_{\nu\lambda}F^2{\rm Tr}(L+A_L)^2
\label{T2}
\ee
All the contributions from the Wess-Zumino term are in the anomalous current $\delta \Gamma/\delta A$. 
(\ref{T2}) is gauge invariant under both a rotation of $U$ and $A_L$. The canonical stress
tensor obeys the equation of motion

\be
\partial^\nu\tilde{T}_{\nu\lambda}=-\partial_\lambda^A\,{\cal L}
\label{T3}
\ee
with ${\cal L}$ the Lagrangian density associated to (\ref{WITTEN}) and $\partial^A$ acting only on the 
$A_L$ fields in ${\cal L}$.  Combining (\ref{T1}) and (\ref{T3}) yields 

\be
\partial^\nu T_{\nu\lambda}=
{\rm Tr}\left(F_{L\nu\lambda}\frac{\delta \Gamma}{\delta A_{L\nu}}\right)
+{\rm Tr}\left(A_{L\lambda}\nabla_\nu\frac{\delta \Gamma}{\delta A_{L\nu}}\right)
\label{T4}
\ee
For non-anomalous left currents, the second contribution in (\ref{T4}) vanishes on-shell,
leaving a gauge-invariant contribution solely from the Lorentz force. For anomalous currents,
this is not the case for an open system.

We note that the gauge-variant contribution in (\ref{T1}) drops locally even in the anomalous case if we were to 
close the system by making $A_L$ dynamical through the addition of $\Gamma^A=-\int {\rm Tr}F_L^2/2$. It also
drops globally by gauge-invariance through

\be
\left<\partial^\nu T_{\nu\lambda}\right>=
{\rm Tr}\left(F_{L\nu\lambda}\left<\frac{\delta \Gamma}{\delta A_{L\nu}}\right>\right)=
{\rm Tr}\left(F_{L\nu\lambda}\,\left<J_{L\nu}\right>\right)
\label{T4X}
\ee
where the expectation value is carried using (\ref{WITTEN}) as the weight in the superfluid ground state,

\be
\left<{\cal O}\right>=\left( \int \, dU\,{\cal O}[U, A_L]\,e^{-i\Gamma [U, A_L]}\right)\,\,
\left(\int \, dU\,e^{-i\Gamma [U, A_L]}\right)^{-1}
\ee
 The last
relation in (\ref{T4X}) involves the gauge invariant but anomalous current. (\ref{T4X}) will be assumed in the 
superfluid state.


\section{Mean-Field Approximation}

To analyze the anomalous current along with the symmetric energy momentum, we define 
the chiral field as $U=e^{i\pi_L}$ with $\pi_L=\pi_L^a\,T^a$  a generic SU(N) valued field.  
In the superfluid state, $\pi_L$ plays the role of the phonons which are excited either 
quantum mechanically or through temperature.  For convenience we also define

\be
2{\rm Tr}T^a(L+A)=ie^{ab}(\pi_L) \,\left(d\pi_L^b+(e^{-1}A_L)^b\right)\equiv ie^{ab}(\pi_L)\,\Pi^b
\label{DEF}
\ee
after using $A_L=iT^aA_L^a$. 
At tree level in the phonon fluctuations with $\left<e^{ab}(\pi_L)\right>\approx \delta^{ab}$, (\ref{CUR2}) and (\ref{T2}) 
are tied by the equation of motion (\ref{T4X}),
with  the bare fields

\begin{eqnarray}
T_{\nu\lambda}\approx&&F^2\,
\Pi_\nu^a\Pi^a_\lambda-\frac 12 g_{\nu\lambda}\,\Pi^a_\alpha\Pi^{a\alpha}\nonumber\\
J_{L}^a\approx&&F^2\,\Pi^a+\frac C4 d^{abc}\,\Pi^{b}d\Pi^{c}
-\frac C4 d^{abc}\Pi^{b}\,dA_L^{c}
\end{eqnarray}

To go beyond tree level in the phonon fluctuations we
need to expand the vierbeins $e^{ab}(\pi _L)$ in $\pi_L$. We will use the
mean-field or tadpole approximation as in~\cite{CHIRAL}.
In the leading order ${\cal O}(\pi_L^4)$

\begin{eqnarray}
&&\left<e^{ab}(\pi_L)\right>\approx \delta^{ab}\left(1-\frac{N\pi_L^2}{6}\right)\nonumber\\
&&\left<e^{bB}(\pi_L)e^{cC}(\pi_L)\right>\approx 
\delta^{bB}\delta^{cC}\left(1-\frac{N\pi_L^2}{12}\right)
\label{MEAN}
\end{eqnarray}
with 

\be
\pi_L^2\equiv\left<\pi_L^2\right> =\frac 1{F^2}\sum_k\frac 1{2\omega_k}
\label{SUM}
\ee
Again all expectations values in the superfluid ground state are carried using (\ref{WITTEN}) as a weight.
At zero temperature, the sum is cutoff by the highest phonon or Debye frequency $\omega_D$ 
(scale of validity of the effective description), while at finite temperature the sum is multiplied by $(1+2\,n_k)$ with $n_k$ the thermal 
phonon distribution.  As noted earlier,  $\omega_k=k$ for relativistic phonons in the tadpole approximation. The phonons
are non-luminal beyond this approximation. The expectation values can be carried to higher order in $\pi_L^2$ through a further expansion
of the vierbeins which amounts to a tadpole resummation. This point will not be pursued here. 

With the above in mind, the energy-momentum tensor now reads

\be
T_{\nu\lambda}\approx \left(F^2\,\Pi_\nu^a\Pi_\lambda^a-\frac 12 {F^2}\,g_{\nu\lambda}\Pi_\alpha^a\Pi_\alpha^a\right)\,
\left(1-\frac {N\pi_L^2}{12}\right)
\label{TA}
\ee
while the gauge-invariant current is

\be
J_{L}^a\approx F^2\,\Pi^a\,\left(1-\frac{N\pi_L^2}6\right)
+\frac C4\,d^{abc}\Pi^b\,d\Pi^c\left(1-\frac{5N\pi_L^2}{24}\right)
-\frac {C}4\,d^{abc}\,\Pi^b\,dA_L^c\,\left(1-\frac {N\pi_L^2}6\right)
\label{JA}
\ee
with the form notation implied. 
In the tadpole approximation both $F$ and $\Pi$ renormalize.  To 
order ${\cal O}(\pi_L^4)$ the renormalizations are

\begin{eqnarray}
F_r\equiv&&F\left(1-\frac{N\pi_L^2}8\right)\nonumber\\
\Pi_r\equiv&&\Pi\left(1+\frac{N\pi_L^2}{12}\right)
\label{REN}
\end{eqnarray}
Inserting (\ref{REN}) in (\ref{TA}) and (\ref{JA}) yield the renormalized energy-momentum tensor

\be
T_{\nu\lambda}\approx F_r^2\,\Pi_{r\nu}^a\Pi_{r\lambda}^a-\frac 12 {F^2_r}\,g_{\nu\lambda}\Pi_{r\alpha}^a\Pi_{r\alpha}^a\,
\label{TAR}
\ee
and renormalized anomalous current

\be
J_{L}^a\approx F_r^2\,\Pi_r^a\,+\frac C4\,d^{abc}\Pi_r^b\,d\Pi_r^c\,\left(1-\frac{3N\pi_L^2}8\right)
-\frac {C}4\,d^{abc}\,\Pi_r^b\,dA_L^c\,\left(1-\frac {N\pi_L^2}4\right)
\label{JAR}
\ee
As expected both the energy-momentum tensor and the normal contribution to $J_L$ are 
renormalized to their tree level contribution through the redefinitions (\ref{REN}). The anomalous
contributions in $J_L$ are not. We will revisit this point below.

\section{Hydrodynamics}

We now identify the expectation value of the renormalized phonon field $\Pi_r^a$ with the local superfluid 4-velocity $v$
through the Cartan chemical potentials $\mu^a$ for the left charges in the absence of the external
field (conserved left currents),

\be
\left<\Pi_r^a(x)\right>\approx \mu^a\,v(x)
\ee
We observe that this identification is in general irrotational, and similar to

\be
J=|\psi^\dagger\psi|\,\frac{(d\phi-eA/c)}m=n\,v
\ee
for the normal current contribution to a charged  and non-relativistic
U(1) superfluid state with wavefunction $\psi=|\psi|e^{i\phi}$. A similar
observation was also noted in~\cite{SON} for the U(1) relativistic 
phonons in dense QCD with $eA/c=\mu$ as the baryon chemical potential.

The normal contribution (first term) in (\ref{JAR}) implies

\be
F_r^2=\frac{n^a}{\mu^a}=\frac{\epsilon+p}{(N-1)\mu^2}
\label{FPI}
\ee
where the last equality follows from the canonical form of the energy-momentum tensor
as in (\ref{TFLUID1}) below, which is also the first law of thermodynamics at zero temperature.
(\ref{FPI}) implies left flavor symmetry in the chiral superfluid or $\mu^a=\mu$ for all Cartan labels $a=1,..,(N-1)$.  
This condition can be
relaxed in the chiral superfluid by explicitly breaking chiral symmetry and rediagonalizing the
chiral phonons each with a different $F$. This notwithstanding, and using (\ref{SUM}) we note that

\be
\frac {N\pi_L^2}{2}=\frac{\mu^2}{\epsilon + p}\left( {N(N-1)}\sum_k\frac 1{2\omega_k}\right)=
\frac {\mu^2}{\epsilon+p}\,\frac{n_{ph}}{\omega_D}
\ee
where the brackets counts the density of phonons $n_{ph}$ per Debye frequency $\omega_D$
at zero temperature. Following Debye's idea for counting phonons in solids, we fix the density of
phonons to match the number of effective degrees of freedom  at the Debye cutoff $\omega_D=\mu$.
Specifically, $n_{ph}=(N-1)n$ with $n^a=n$ for all Cartan charges.  Thus

\be
\frac{N\pi_L^2}2=\frac{n^a\mu^a}{\epsilon +p}
\label{RATIO}
\ee
Inserting (\ref{RATIO}) and the superfluid identification in the energy-momentum tensor (\ref{TAR})
yield

\be
\left<T_{\nu\lambda}\right>\approx(\epsilon+p)\left(v_\nu\,v_\lambda-\frac 12\,g_{\nu\lambda}\,v_\alpha\,v^\alpha\right) +
\frac {\Delta}4 g_{\nu\lambda}
\label{TFLUID1}
\ee
after adding the superfluid ground state energy per volume $\Delta$. Inserting (\ref{RATIO}) in the anomalous current
(\ref{JAR}) yields

\be
\left<J_{L}^a\right> \approx n^a\,v+\frac{C}4 \,d^{abc}\,\mu^b\,\mu^c\, v\,dv\,\left(1-\frac 34  \frac{n^e\mu^e}{\epsilon+p}\right)
-\frac {C}4\,d^{abc}\,\mu^b\,v\,dA_L^c\,\left(1-\frac 12\,\frac{n^e\mu^e}{\epsilon+p}\right)
\label{JFLUID}
\ee

Our  mean-field or leading tadpole approximation result (\ref{JFLUID}) for the left Cartan current 
is overall similar to a recent result established in~\cite{SON1} for the axial vector Cartan current. 
Indeed, for the latter we expect $C/2\rightarrow C$ as it sums left plus right, and also $dA_L/2\rightarrow dA$
since $A=(A_R+A_L)/2$ for an external Cartan vector field.  The overall sign of $dA$ is conventional.
With this in mind, we note a difference in the first tadpole 
contribution with $3/4$ in our case versus $2/3$ in~\cite{SON1}. It maybe traced back to
the fact that in our case the equations of motion are field equations following from varying say 
$\Pi_r\approx \mu\ v$ and
not $\mu$ and $v$ individually as in the constitutive analysis in~\cite{SON1}.  Our analysis is fully quantum 
mechanical beyond the tadpole approximation. For instance, at zero temperature the tadpole corrections in 
(\ref{JFLUID}) are not small, suggesting that all tadpoles in the vierbeins in (\ref{MEAN}) need to be resummed.

Finally, the tadpole or mean-field corrections to the anomalous flavor current in the superfluid state appear at 
odd with the Adler-Bardeen's non-renormalization theorem of the flavor anomaly in QCD both at finite temperature
and density~\cite{BARDEEN}.  The origin of this discrepancy is again the approximation
(\ref{MEAN}) that leaves out tadpole-like corrections originating from the anomalous vertices. We have checked
that when these corrections are taken into account they result in only the renormalization of $F$ and $\Pi$ as in 
(\ref{REN}). They are also equation of motion preserving for the fully renormalized energy-momentum tensor and anomalous
current. However, this tadpole approximation with vertex renormalizations is quantum mechanical in nature and 
goes beyond the realm of a constitutive analysis.

\section{Conclusions}

We have shown how global non-Abelian flavor anomalies can be translated to the global constitutive
currents from the microscopic currents using the anomaly equation in a superfluid state.  In the mean-field
or tadpole approximation, the gauge-invariant and symmetric part of
the energy momentum tensor renormalizes to an ideal fluid form. The gauge-invariant and anomalous 
flavor current renormalizes to a normal and ideal fluid contribution plus anomalous corrections. The latter
are similar to a recent result established recently using the constitutive equations and thermodynamics
\cite{SON1}.  Our arguments trace the origin of the anomalous contributions to the Wess-Zumino-Witten term. The
Adler-Bardeen non-renormalization theorem shows that these mean-field corrections eventually renormalize away
when quantum vertex corrections are included.

Flavor anomalies play an intriguing role in models of QCD at high density and low temperature,
and should be of relevance to a number of time-odd superfluid effects in QCD at high density~\cite{CFL} whether
in the form of cold hadronic neutron stars with magnetic fields, dense quark superfluidity in the presence of the
twisted U(1) photon or even perhaps a heavy ion collision with U$_A$(1) violating fields~\cite{DIMA}. 
Also these flavor anomalies naturally arise in current holographic models of QCD or their 
variants at finite density  and low temperature in the presence of external flavor fields and in the long wavelength
limit.

\section{Acknowledgments}
We thank Edward Shuryak for discussions.
This work was supported in part by US-DOE grants
DE-FG02-88ER40388 and DE-FG03-97ER4014.

\end{document}